\documentclass{ws-procs9x6}
\newcommand{\beq}{\begin{equation}}
\newcommand{\eeq}{\end{equation}}

\newcommand{\lsim}{\mathrel{\mathop{\kern 0pt \rlap
{\raise.2ex\hbox{$<$}}}
\lower.9ex\hbox{\kern-.190em $\sim$}}}
\newcommand{\gsim}{\mathrel{\mathop{\kern 0pt \rlap
{\raise.2ex\hbox{$>$}}}
\lower.9ex\hbox{\kern-.190em $\sim$}}}

\begin{document}

\title{NEUTRALINO CLUMPS AND COSMIC RAYS}

\author{P. SALATI$^*$}

\address{
LAPTH, 9 Chemin de Bellevue, BP110,\\
F--74941 Annecy--le--Vieux Cedex, France\\
$^*$E-mail: salati@lapp.in2p3.fr\\
http://lappweb.in2p3.fr/~salati/}
\address{{\large Lapth-Conf-1172/07}
\vskip 0.1cm
Proceedings of the
{\sl 6th International Workshop on the Identification of Dark Matter},
held in Rhodes Island, Greece, September 11--16, 2006}
%%%%%%%%%%%%%%%%%%%%%%%%%%%%%%%%%%%%%%%%%%%%%%%%%%%%%%%%%%%%%%%%%%%%%%%%%%%
%%%%%%%%%%%%%%%%%%%%%%%%%%%%%%%%%%%%%%%%%%%%%%%%%%%%%%%%%%%%%%%%%%%%%%%%%%%
\begin{abstract}
The halo of the Miky Way might contain numerous and dense substructures inside
which the putative weakly interacting massive particles (suggested as the main
constituent of the astronomical dark matter) would produce a stronger annihilation
signal than in the smooth regions. The closer the nearest clump, the larger the
positron and antiproton cosmic ray fluxes at the Earth. But the actual distribution
of these substructures is not known. The predictions on the antimatter yields at
%the Earth are therefore affected by some sort of cosmic variance whose analysis
the Earth are therefore affected by a kind of cosmic variance whose analysis
is the subject of this contribution. The statistical tools to achieve that goal
are presented and Monte Carlo simulations are compared to analytic results.
\end{abstract}
\bodymatter
%%%%%%%%%%%%%%%%%%%%%%%%%%%%%%%%%%%%%%%%%%%%%%%%%%%%%%%%%%%%%%%%%%%%%%%%%%%
%%%%%%%%%%%%%%%%%%%%%%%%%%%%%%%%%%%%%%%%%%%%%%%%%%%%%%%%%%%%%%%%%%%%%%%%%%%
\section{Motivations}

%
%FFFFFFFFFFFFFFFFFFFFFFFFFFFFFFFFFFFFFFFFFFFFFFFFFFFFFFFFFFFFFFFFFFFFFFFFFF
\begin{sidewaysfigure}
\begin{center}
\centerline{
\psfig{file=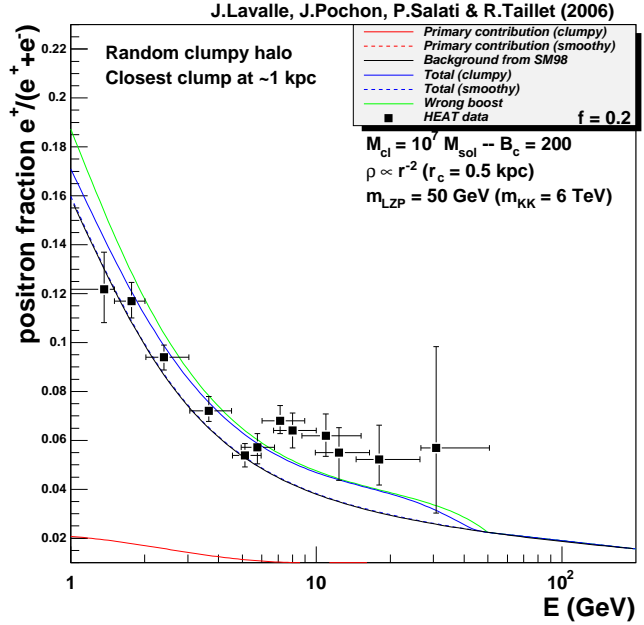,width=3.70in}
\psfig{file=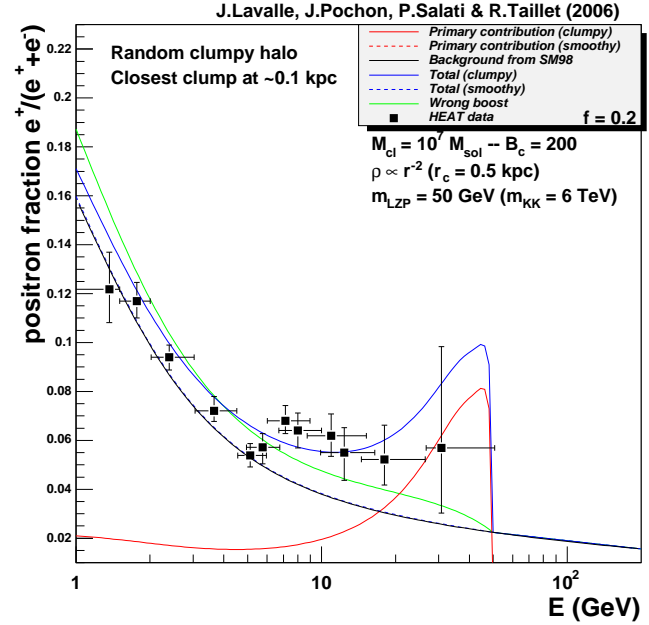,width=3.70in}}
\end{center}
\caption{
The positron fraction is derived for a 50 GeV Kaluza--Klein inspired particle
\cite{agashe_servant_04} and compared to the HEAT excess \cite{heat2} which
cannot be explained by the pure conventional background \cite{sm98} in black.
An isothermal profile has been assumed for the dark matter halo of the Milky Way,
a fraction $f = 0.2$ of which lies in clumps with mass $10^{7}$ M$_{\odot}$
and intrinsic boost $B_{c} = 200$. Two random realizations of that clumpy DM
halo are featured in blue. The distance of the closest substructure has been set
equal to 1 kpc in the left panel and decreased to 0.1 kpc in the right panel.
The green curve corresponds to the traditional and wrong shift by a factor of
$f \times B_{c} = 40$ of the positron spectrum which a completely smooth DM halo
would yield.
}
\label{Fig:jack_pot}
\end{sidewaysfigure}
%FFFFFFFFFFFFFFFFFFFFFFFFFFFFFFFFFFFFFFFFFFFFFFFFFFFFFFFFFFFFFFFFFFFFFFFFFF
%

The universe contains a significant fraction of invisible and non--baryonic
matter \cite{Spergel:2006hy} whose nature is still unresolved. This dark matter
could be made of weakly interacting massive particles \cite{Bertone:2004pz},
such as the supersymmetric neutralinos, whose annihilations inside the galactic
halo might produce a sizeable yield of positrons and antiprotons. These rare
antimatter cosmic rays will be detected with improved accuracy by the forthcoming
experiments \cite{AMS,PAM}.
Numerical simulations of structure formation indicate that dark matter (DM)
is clumpy on small scales \cite{Diemand:2005vz}. Because neutralinos are more
tightly packed inside the DM substructures, their mutual annihilation is enhanced.
The resulting increase of the associated antimatter signals at the Earth has been
so far accounted for by an overall boost factor and by shifting upwards the cosmic
ray fluxes obtained in the case of a smooth distribution of DM particles.

This approach is too simple. If a substructure were to lie in our
immediate vicinity, we would observe a significantly distorted spectrum
as illustrated in \fref{Fig:jack_pot} for a 50 GeV LZP species. 
The question then naturally arises to know if such a possibility is probable
or very rare. Should we know the exact location of each DM clump, the
positron and antiproton spectra would be derived exactly. On the contrary,
they are affected by a kind of cosmic variance because an infinite number of
different halo realizations are possible. The theoretical predictions on the
detectable fluxes at the Earth must take into account that lack of knowledge.

We have therefore built specific tools \cite{Lavalle:2006vb}
to address this issue. We will concentrate
here on the pedagogical example where a positron line is produced through the mutual
annihilations of the DM particles $\chi$ into $e^{+} e^{-}$ pairs. This scenario
is plausible in the framework of Kaluza--Klein inspired theories.
In the absence of any substructure, the DM distribution $\rho_{s}$ is smooth
and the resulting flux is given by
\beq
\phi_{s}(E) = {\mathcal S} \,
{\displaystyle \int}_{\rm DM \, halo} \!\! G(\mathbf{x}) \,
\left\{
{\displaystyle \frac{\rho_{s}(\mathbf{x})}{\rho_{0}}}
\right\}^{2} \, d^{3} \mathbf{x} \;\; ,
\eeq
where the factor ${\mathcal S}$ is defined as
\begin{equation}
\mathcal{S} = \frac{1}{8 \pi} \, v_{e^+}(E) \,
\langle \sigma_{\rm ann}
\left( \chi \, \chi \to e^+ \, e^{-} \right) \, v \rangle \,
\left( \frac{\rho_{0}}{m_{\chi}} \right)^2 \;\; .
\end{equation}
The probability for a positron injected at $\mathbf{x}$ to propagate towards
the Earth which it reaches with the degraded energy $E \leq E_{S}$ is described
by the Green function $G(\mathbf{x})$.
In the presence of clumps, the DM distribution is given by the superposition
$\rho = \rho'_{s} + \delta \rho$ where a smooth component $\rho'_{s}$
is still present while $\delta \rho$ accounts for the substructures.
The propagator $G$ has already been discussed in the literature \cite{Baltz:1998xv}.
Positrons loose energy through synchrotron radiation and inverse Compton scattering
on stellar light and on the CMB. The farther they originate, the smaller their energy
$E$ at the Earth for a fixed value of $E_{S}$. Positrons that are detected at the
energy $E$ have been produced within a typical range $\lambda_{\rm D}$ which
decreases as $E$ increases towards $E_{S}$ but is nevertheless much larger than
the size of the DM substructures. The positron flux becomes
\beq
\phi(E) = \phi'_{s} +
\left( \phi_{r} = {\displaystyle \sum_{i}} \; \varphi_{i} \right) \;\; ,
\eeq
where the contribution from the ith clump is
$\varphi_{i} = {\mathcal S} \times G(\mathbf{x}_{i}) \times \xi_{i}$. That
minihalo produces as many positrons as if the entire volume
\beq
\xi_{i} \equiv {\displaystyle \frac{B_{i} \, M_{i}}{\rho_{0}}} =
{\displaystyle \int}_{\rm ith \, clump} \!
\left\{
{\displaystyle \frac{\delta \rho(\mathbf{x})}{\rho_{0}}}
\right\}^{2} \, d^{3} \mathbf{x}
\eeq
were filled with the density $\rho_{0}$. The volume $\xi_{i}$ can also be
expressed as a function of the substructure mass $M_{i}$ and {\bf intrinsic}
boost $B_{i}$.

An infinite set of halo realizations must be taken into account, each of which
produces a different flux $\phi_{r}$. The boost factor
$B = \phi / \phi_{s}$ {\bf at the Earth} is not unique and must be treated as
a random variable. We present in the next section a procedure to determine the
statistical law according to which $B$ is distributed and will show that its
average value and variance depend on the energy $E$.

%%%%%%%%%%%%%%%%%%%%%%%%%%%%%%%%%%%%%%%%%%%%%%%%%%%%%%%%%%%%%%%%%%%%%%%%%%%
%%%%%%%%%%%%%%%%%%%%%%%%%%%%%%%%%%%%%%%%%%%%%%%%%%%%%%%%%%%%%%%%%%%%%%%%%%%
\section{Computing the Odds of the Galactic Lottery}

The statistical properties of the random flux $\phi_{r}$ and of the associated
boost factor $B$ have been thoroughly investigated in a recent analysis
\cite{Lavalle:2006vb}. I will just point out its salient features and summarize
the hypotheses on which it is based.

\noindent {\bf (i)}
To simplify the discussion and without loss of generality, we assume identical
clumps with mass $M_{c}$ and intrinsic boost $B_{c}$. The random flux $\phi_{r}$
simplifies into
\beq
\phi_{r} = {\mathcal S} \times \xi_{c} \times {\displaystyle \sum_{i}} \; G_{i}
\;\; ,
\eeq
where the effective volume $\xi_{c} = {B_{c} \, M_{c}}/{\rho_{0}}$ is the same
for each minihalo.

\noindent {\bf (ii)}
The actual distribution of DM substructures is one particular realization to
be taken from the statistical ensemble made up by all the possible {\bf random}
distributions. The flux contributed by the clumps and the boost factor must be
averaged on that infinite set to yield $\langle \phi_{r} \rangle$ and
$B_{\rm eff} = \langle B = ({\phi}/{\phi_{s}}) \rangle$ whereas the variance
is defined as
\beq
\sigma_{r}^{2} =
\langle \phi_{r}^{2} \rangle \, - \, \langle \phi_{r} \rangle^{2}
\;\;\;\;\;{\rm and}\;\;\;\;\;
\sigma_{B} = {\sigma_{r}}/{\phi_{s}} \;\; .
\eeq

\noindent {\bf (iii)}
Clumps are distributed independently of each other. This is a strong assumption
that generally holds because the substructure correlation length is smaller than
the propagation range $\lambda_{\rm D}$. We just need then to determine how a single
clump is distributed inside the Milky Way halo in order to derive the statistical
properties of an entire population of $N_{H}$ such substructures. If $\varphi$
denotes the contribution of a single minihalo, we get
\beq
\left\langle \phi_{r} \right\rangle = N_{H}
\left\langle \varphi \right\rangle
\;\;\;{\rm and}\;\;\;
\sigma_{r}^{2} = N_{H} \, \sigma^{2} = N_{H}
\left\{
\langle \varphi^{2} \rangle \, - \,
\langle \varphi \rangle^{2}
\right\} \;\; .
\eeq

%
%FFFFFFFFFFFFFFFFFFFFFFFFFFFFFFFFFFFFFFFFFFFFFFFFFFFFFFFFFFFFFFFFFFFFFFFFFF
\begin{figure}
\begin{center}
\psfig{file=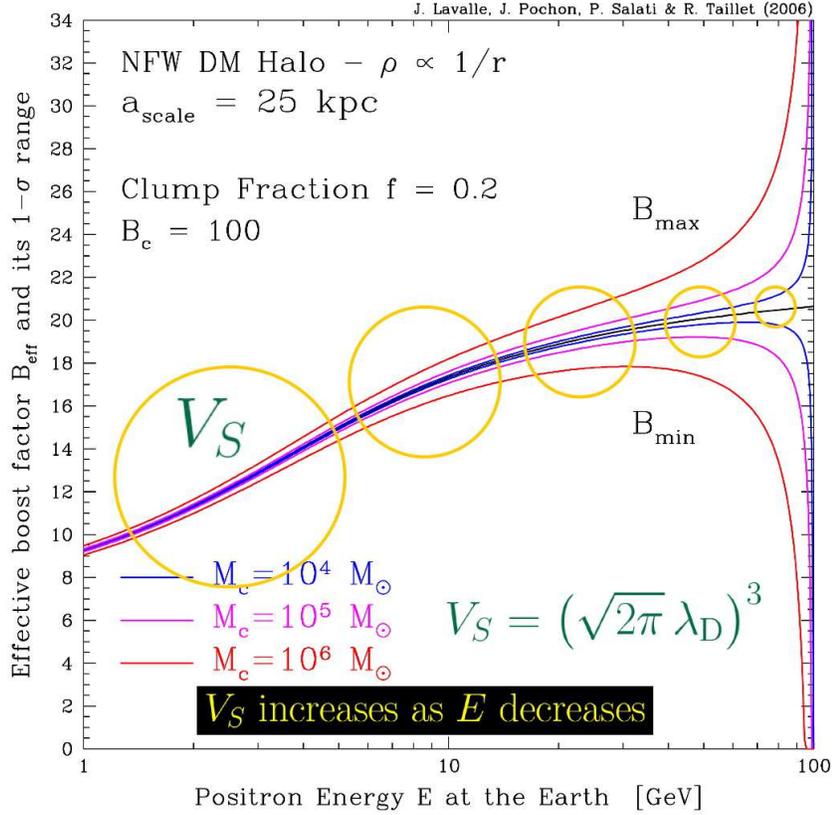,width=4.5in}
\end{center}
\vskip -0.2cm
\caption{
Positrons are injected with the energy $E_{S} = 100$ GeV and detected
at the energy $E$. The effective boost factor $B_{\rm eff}$ (black line)
accounts for the average enhancement of the positron signal resulting from
DM clumpiness. Although it has been so far considered to behave as a constant,
the boost actually depends on the energy.
Furthermore, as $E$ approaches $E_{S}$, the boost variance $\sigma_{B}$
increases significantly. The region from which the positrons detected at
the Earth originate shrinks and the number $N_{S}$ of clumps which it
contains decreases.
}
\label{Fig:boost}
\end{figure}
%FFFFFFFFFFFFFFFFFFFFFFFFFFFFFFFFFFFFFFFFFFFFFFFFFFFFFFFFFFFFFFFFFFFFFFFFFF
%

\noindent {\bf (iv)}
The set of the random distributions of one single clump inside the Milky
Way halo ${\mathcal D}_{H}$ makes up the statistical ensemble ${\mathcal T}$
which we eventually need to consider. An event from that ensemble consists
in a clump located at position $\mathbf{x}$ within the elementary volume
$d^{3} \mathbf{x}$ and is weighted by the probability
$p(\mathbf{x}) \, d^{3} \mathbf{x}$. The spatial distribution $p(\mathbf{x})$
and the flux density of probability ${\mathcal P} \left( \varphi \right)$
are related by
\beq
{\mathcal P} \left( \varphi \right) \, d\varphi \; = \; dP \; = \;
{\displaystyle \int_{{\mathcal D}_{\varphi}}} \,
p(\mathbf{x}) \, d^{3} \mathbf{x} \;\; ,
\eeq
where ${\mathcal D}_{\varphi}$ is the region of the halo in which a clump
must lie in order to contribute a flux $\varphi$ at the Earth. We readily
conclude that any function $\mathcal{F}$ of the flux $\varphi$
(for instance $\varphi$ itself or its square ${\varphi}^{2}$)
is given on average by
\beq
\langle \mathcal{F} \rangle =
{\displaystyle \int} \, \mathcal{F} \! \left( \varphi \right) \,
{\mathcal P} \left( \varphi \right) \, d\varphi \; = \;
{\displaystyle \int_{{\mathcal D}_{H}}} \,
\mathcal{F} \left\{ \varphi \left( \mathbf{x} \right) \right\} \,
p(\mathbf{x}) \, d^{3} \mathbf{x} \;\; .
\eeq
Once $p(\mathbf{x})$ is known, $B_{\rm eff}$ and $\sigma_{B}$ can be derived.

%%%%%%%%%%%%%%%%%%%%%%%%%%%%%%%%%%%%%%%%%%%%%%%%%%%%%%%%%%%%%%%%%%%%%%%%%%%
%%%%%%%%%%%%%%%%%%%%%%%%%%%%%%%%%%%%%%%%%%%%%%%%%%%%%%%%%%%%%%%%%%%%%%%%%%%
\section{Results and Perspectives}

\begin{figure}
\begin{center}
\psfig{file=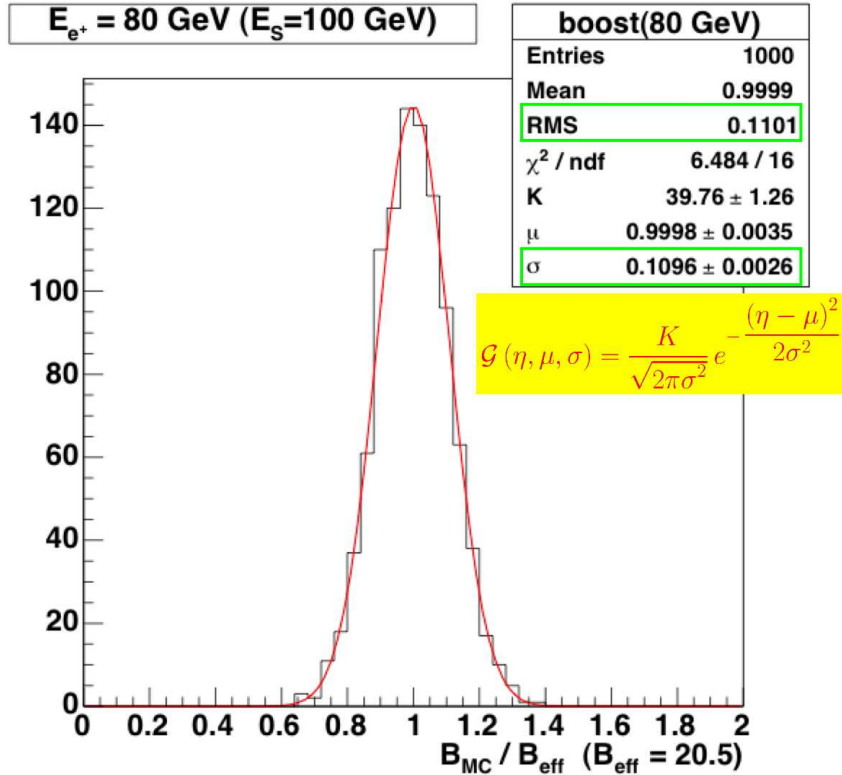,width=4.5in}
\end{center}
\vskip -0.2cm
\caption{
In this Monte--Carlo simulation, clumps contribute a fraction $f = 0.2$
to the mass of the Milky Way DM halo (assumed to follow a NFW profile
with a scale radius of 25 kpc). Each substructure has a mass of
$10^{5}$ M$_{\odot}$. A thousand different realizations of the DM halo
have been generated here, each of them involving 271,488 clumps. The
positron injection energy is $E_{S} = 100$ GeV. The number of realizations
has been plotted as a function of the boost ratio $\eta = {B}/{B_{\rm eff}}$.
The rms value of 0.1101 yielded by the Monte--Carlo for $\eta$ is in excellent
agreement with the anaytic value
$\sigma_{\eta} = {\sigma_{B}}/B_{\rm eff} = 0.1097$.
}
\label{Fig:large_NS}
\end{figure}

We have applied that procedure in the case where
$\rho'_{s} \equiv (1 - f) \times \rho_{s}$ and have finally assumed that
the clumps trace the smooth DM distribution (which is not the case in
general) by imposing that $p(\mathbf{x}) \propto \rho_{s}(\mathbf{x})$.
The main results are summarized in \fref{Fig:boost} and \ref{Fig:large_NS}.

To commence, the average boost $B_{\rm eff}$ depends on the energy. As $E$
decreases, the part of the halo which contributes effectively to the positron
signal at the Earth grows. Regions located close to the galactic center come
into play, with a much larger density $\rho_{s}$ than in the solar
neighborhood. At fixed intrinsic boost $B_{c}$, neutralino annihilation
inside clumps is relatively less enhanced in these dense regions than in
our vicinity, hence the behaviour exhibited in \fref{Fig:boost}.

We also find that the boost variance $\sigma_{B}$ increases significantly
as $E$ approaches $E_{S}$. The above mentioned domain of the halo, inside
which the positrons detected at the Earth with the energy $E$ have been
produced, has a typical volume
$V_{S} = (\sqrt{2 \pi} \, \lambda_{\rm D})^{3}$ which shrinks
as $E$ gets near to $E_{S}$. The number $N_{S}$ of the substructures which
it contains falls down and the boost variance increases because
\beq
{\displaystyle \frac{\sigma_{r}}{\langle \phi_{r} \rangle}} \simeq
{\displaystyle \frac{\sigma_{B}}{B_{\rm eff}}} \simeq
{\displaystyle \frac{1}{\sqrt{\displaystyle N_{S}}}} \;\; .
\eeq

Finally, in the large $N_{S}$ regime, the boost distribution follows a
Maxwellian law as is featured in \fref{Fig:large_NS}. This result may be
derived with the help of the central limit theorem. In the opposite regime
where $N_{S} \leq 1$, the boost distribution is driven by the product
$N_{H} \times {\mathcal P} \left( \varphi \right)$ where the density of
probability ${\mathcal P} \left( \varphi \right)$ for the flux generated
by a single clump comes into play.

The method outlined here can be applied in particular to the intermediate mass
black hole scenario \cite{Bertone:2005xz}. The spatial distribution $p(\mathbf{x})$
no longer traces the smooth density $\rho_{s}$. An effective boost $B_{\rm eff}$
of $\sim 2$ to $3 \times 10^{3}$ is found \cite{brun06} for both antiprotons
and positrons with a large variance.

%
%FFFFFFFFFFFFFFFFFFFFFFFFFFFFFFFFFFFFFFFFFFFFFFFFFFFFFFFFFFFFFFFFFFFFFFFFFF
%FFFFFFFFFFFFFFFFFFFFFFFFFFFFFFFFFFFFFFFFFFFFFFFFFFFFFFFFFFFFFFFFFFFFFFFFFF
%

\vskip -0.2cm
%%%%%%%%%%%%%%%%%%%%%%%%%%%%%%%%%%%%%%%%%%%%%%%%%%%%%%%%%%%%%%%%%%%%%%%%%%%
%%%%%%%%%%%%%%%%%%%%%%%%%%%%%%%%%%%%%%%%%%%%%%%%%%%%%%%%%%%%%%%%%%%%%%%%%%%
\section*{Acknowledgments}
I would like to thank the organizers for the warm and friendly atmosphere
of this exciting and inspiring conference as well as for their financial support.
This work could not have been performed without the help of the french programme
national de cosmologie PNC.

%%%%%%%%%%%%%%%%%%%%%%%%%%%%%%%%%%%%%%%%%%%%%%%%%%%%%%%%%%%%%%%%%%%%%%%%%%%
%%%%%%%%%%%%%%%%%%%%%%%%%%%%%%%%%%%%%%%%%%%%%%%%%%%%%%%%%%%%%%%%%%%%%%%%%%%

%
%\bibliographystyle{ws-procs9x6}
%\bibliography{salati_rhodes_biblio}

\begin{thebibliography}{99}

\bibitem{Spergel:2006hy}
D.~N. Spergel {\em et~al.}, {\tt astro-ph/0603449} (2006).

\bibitem{Bertone:2004pz}
G.~Bertone, D.~Hooper and J.~Silk, {\em Phys. Rept.} {\bf 405}, 279 (2005).

\bibitem{AMS}
F.~Barao  [AMS-02 Collaboration],
%``AMS: Alpha Magnetic Spectrometer on the International Space Station,''
{\em Nucl. Instrum. Meth. A} {\bf 535}, 134 (2004).

\bibitem{PAM}
P.~Picozza {\em et~al.},
%``PAMELA: A payload for antimatter matter exploration and light-nuclei
%astrophysics,''
{\tt astro-ph/0608697} (2006).

\bibitem{Diemand:2005vz}
J.~Diemand, B.~Moore and J.~Stadel, {\em Nature} {\bf 433}, 389 (2005).

\bibitem{agashe_servant_04}
K.~{Agashe} and G.~{Servant}, {\em Phys. Rev. Letters} {\bf 93}, 231805 (2004).

\bibitem{heat2}
S.~{Coutu} {\em et~al.}, {Positron Measurements with the Heat-Pbar Instrument}, in
{\em International Cosmic Ray Conference\/}, 2001.

\bibitem{sm98}
I.~V. {Moskalenko} and A.~W. {Strong}, {\em Astrophys. J.} {\bf 493}, 694 (1998).

\bibitem{Lavalle:2006vb}
J.~Lavalle, J.~Pochon, P.~Salati and R.~Taillet, {\tt astro-ph/0603796}, to be
published in A\&A (2007).

\bibitem{Baltz:1998xv}
E.~A. Baltz and J.~Edsjo, {\em Phys. Rev.} {\bf D59}, p. 023511 (1999).

\bibitem{Bertone:2005xz}
G.~Bertone, A.~R. Zentner and J.~Silk, {\em Phys. Rev.} {\bf D72}, p. 103517 (2005).

\bibitem{brun06}
P.~Brun, G.~Bertone, J.~Lavalle, P.~Salati and R.~Taillet, in preparation (2007).

\end{thebibliography}

%%%%%%%%%%%%%%%%%%%%%%%%%%%%%%%%%%%%%%%%%%%%%%%%%%%%%%%%%%%%%%%%%%%%%%%%%%%
%%%%%%%%%%%%%%%%%%%%%%%%%%%%%%%%%%%%%%%%%%%%%%%%%%%%%%%%%%%%%%%%%%%%%%%%%%%
\end{document}